\documentclass[aps,prl,twocolumn,superscriptaddress,showpacs,floatfix]{revtex4-2}
\usepackage{physics}
\usepackage{graphicx}% Include figure files
\usepackage{dcolumn}% Align table columns on decimal point
\usepackage{bm}% bold math
\usepackage{mhchem}
\usepackage{bbold}
\usepackage{multirow}
\usepackage{subfiles}
\usepackage[colorlinks=true, linkcolor=blue, citecolor=blue, urlcolor=blue]{hyperref}
\usepackage[dvipsnames]{xcolor}

\DeclareUnicodeCharacter{2212}{-}
\usepackage{amsmath}
\usepackage{algorithm}
\usepackage{algpseudocode}
\begin{document}

\preprint{APS/123-QED}

\title{
Symmetry-controlled thermal activation in pyramidal Coulomb clusters:\\ Testing Kramers-Langer theory
}

\author{Akhil Ayyadevara}
%\email{akhilv@rri.res.in}
\affiliation{Raman Research Institute, C. V. Raman Avenue, Sadashivanagar, Bangalore 560080, India}
\author{Anand Prakash}%
\affiliation{Raman Research Institute, C. V. Raman Avenue, Sadashivanagar, Bangalore 560080, India}
\author{Shovan Dutta}
\affiliation{Raman Research Institute, C. V. Raman Avenue, Sadashivanagar, Bangalore 560080, India}
\author{Arun Paramekanti}
\affiliation{Department of Physics, University of Toronto, 60 St. George Street, Toronto, ON, M5S 1A7 Canada}
\author{S. A. Rangwala}
%\email{sarangwala@rri.res.in}
\affiliation{Raman Research Institute, C. V. Raman Avenue, Sadashivanagar, Bangalore 560080, India}

\date{\today}%

\begin{abstract}

Laser-cooled ions confined in electromagnetic traps provide a unique, tunable mesoscopic system where the interplay of the trapping potential, nonlinear Coulomb interactions, and laser-ion scattering generates rich, collective dynamics. In this work, we engineer thermally activated switching between two oppositely oriented, square-pyramidal configurations of five laser-cooled ions in a Paul trap. For identical ions ($^{40}\mathrm{Ca}^{+}$), the inversions proceed via a \textit{Berry pseudo-rotation} mechanism with a low activation barrier, enabled by the permutation symmetry, in contrast to the \textit{umbrella inversion} observed in ammonia. The experimentally measured inversion rates, spanning two orders of magnitude, are accurately captured by the multidimensional Kramers-Langer theory, enabling thermometry of the Doppler-cooled ion cluster at $1.8\pm0.1$ mK. By substituting the apex ion with a heavier isotope ($^{44}\mathrm{Ca}^{+}$), we break the permutation symmetry and observe a suppression of thermally activated inversions. Numerical analysis reveals that this symmetry breaking closes the low-barrier channel, forcing the system to invert through a high-barrier \textit{turnstile rotation}. Thus, we demonstrate a structural analogue of molecular kinetic isotope effects, establishing trapped ions as a versatile platform to explore symmetry-controlled collective dynamics.

\end{abstract}

\maketitle

Rare events|processes that manifest on timescales far exceeding a system's intrinsic dynamics|are observed in diverse physical settings ranging from protein folding and chemical reactions to nucleation and phase transitions. In thermal equilibrium, these events arise when inherent thermal noise drives the system out of a metastable state. A central theme in statistical mechanics is to develop a predictive, system-independent theory for these kinetics \cite{Peters2017}. Kramers provided a foundational framework \cite{Kramers1940, Kramers_review_hanggi} by modelling such transitions as resulting from a Brownian motion in one dimension (1D). Pioneering experiments with single particles in optical traps \cite{McCann1999} and levitated nanoparticles \cite{Rondin2017} have provided remarkable, parameter-free verifications of Kramers' theory based on real-time tracking \cite{Li2010} and effective 1D reaction coordinates \cite{Berezhkovskii2005, Zijlstra2020}. However, complex many-body processes like protein folding, molecular isomerization, and nuclear fission occur in multidimensional energy landscapes, described by the generalized Kramers-Langer (K-L) formalism \cite{Langer1969}. Testing this multidimensional framework requires a mesoscopic system close to thermal equilibrium with a complex and controllable energy landscape. 

Here, we demonstrate that self-organized Coulomb clusters of laser-cooled trapped ions provide an ideal platform for this purpose. By tuning the anisotropy of the confinement, such a cluster can be steered into a configurational bistability \cite{ayyadevara2025, mizukami2025}. The Doppler-cooling lasers provide an effective thermal bath \cite{Javanainen_1980, Javanainen_1986, Morigi_2001}, inducing stochastic photon kicks that can activate the transitions, which can be tracked in real time by measuring the scattered photons. Crucially, as the ions are strongly coupled, local kicks are immediately transferred to all of the ions via collective vibrational modes. Moreover, due to a precise knowledge of the many-body Hamiltonian, these modes and the resulting transition rates from K-L theory can be computed accurately \cite{Landa_2012, Kaufmann_2012, Kiethe_2021, ayyadevara2025}. 

\begin{figure*}[t]
    \centering
    \includegraphics[width=0.99\textwidth]{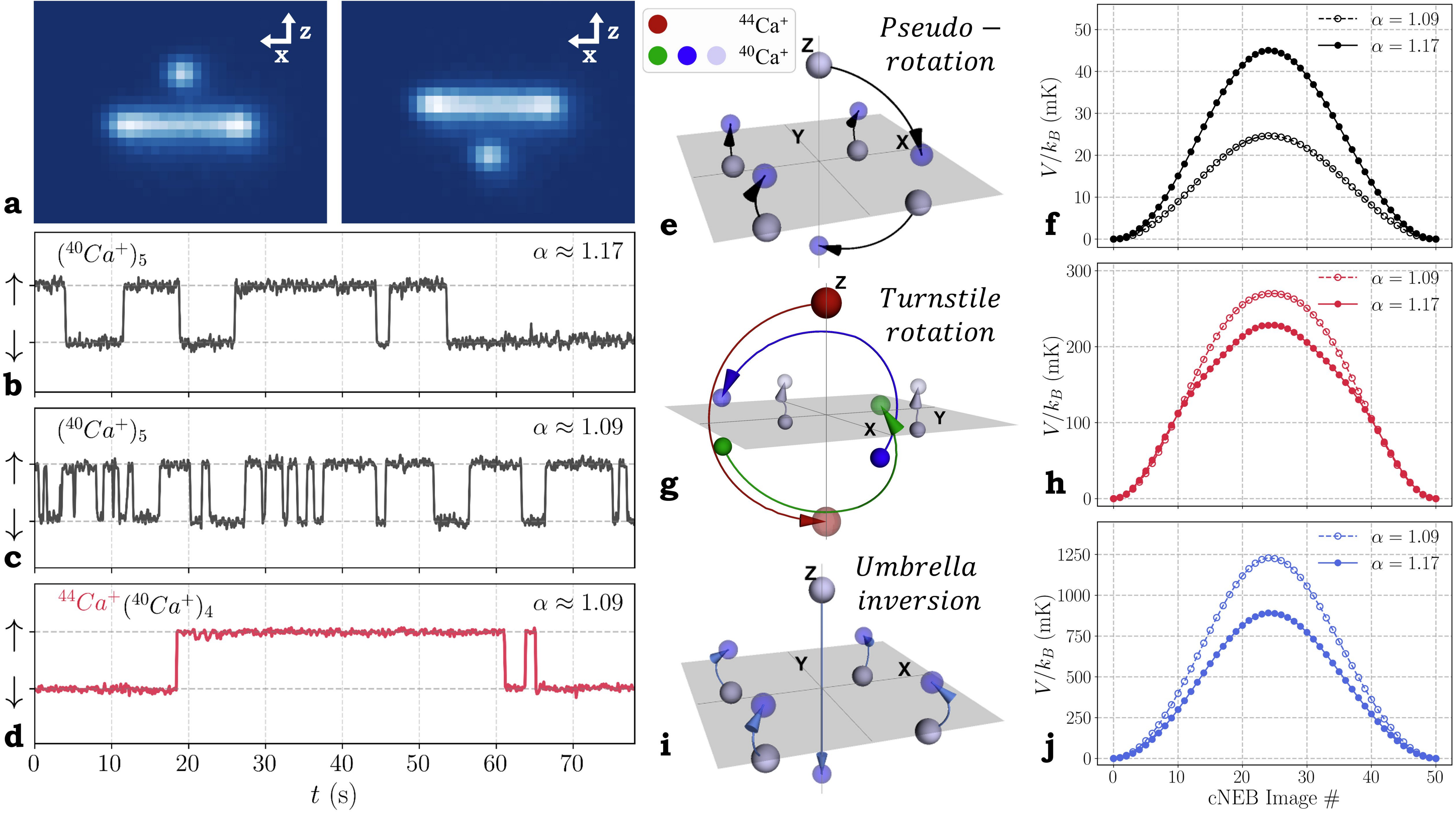}
    \caption{\label{fig1}Inversion mechanisms for a pyramidal ion cluster. (\textbf{a}) Fluorescence images of the two symmetry-broken pyramidal configurations. (\textbf{b, c}) Experimental time traces of the parity-odd octupole moment $\psi_{3,0}$, showing inversions for an identical-ion cluster at two different settings for the trap aspect ratio $\alpha$. (\textbf{d}) Suppression of thermal activation after the apex ion is replaced by a heavier isotope. The rare transitions are triggered by background-gas collisions. (\textbf{e, f}) Inversion pathway for the identical cluster is a low-barrier {\it pseudo-rotation}. (\textbf{g, h}) For the isotope-substituted cluster, the rate-limiting path is a \textit{turnstile rotation} with a much higher activation barrier.  (\textbf{i, j}) The canonical {\it umbrella inversion} involves enormous energy barriers.}
\end{figure*}

We experimentally realize the inversion dynamics of a square-pyramidal cluster of five $^{40}$Ca$^+$ ions [Fig.~\ref{fig1}(a)]. While geometrically analogous to pyramidal molecules like ammonia, the transition pathway of this Coulomb system reveals a richer structural rearrangement. In particular, we demonstrate how the permutation symmetry of the ions opens a low-energy pathway, allowing the system to circumvent the high activation barrier of the canonical umbrella inversion. We harness the tunability of the energy barrier to conduct a parameter-free test of the multidimensional K-L theory. Furthermore, we test the importance of permutation symmetry by substituting the apex with a heavier isotope ($^{44}$Ca$^+$), which turns off the low-energy channel, resulting in a directly observable suppression of the inversions.

{\it \textbf{Pyramidal bistability}}|We trap and laser cool Ca$^+$ ions in an end-cap type radio-frequency (\textit{rf}) trap, as detailed in our previous studies \cite{Anand_oven, ayyadevara2025}. The ions experience a time-averaged harmonic confinement, with an angular frequency $\omega_z$ along the axial direction and degenerate angular frequencies $\omega_x = \omega_y$ along the transverse directions due to cylindrical symmetry of the trap. By superimposing a \textit{dc} voltage onto the electrodes, we can continuously change the aspect ratio of the confinement, $\alpha = \omega_z / \omega_x$; see Supplemental Material (SM) \cite{supp}. The potential energy of $5$ identical ions of mass \textit{m} is given by
\begin{equation}
    V = \sum_{i = 1}^5 \Bigg[ \frac{m \omega_{x}^2}{2} \, \big(x_{i}^2+y_{i}^2+\alpha^{ 2}z_{i}^2 \big) + \sum_{j > i}^5 \frac{k_e e^2}{|\mathbf{r}_{i}- \mathbf{r}_{j}|} \Bigg] ,
\end{equation}
where $\mathbf{r}_i \equiv (x_i, y_i, z_i)$ are the positions of the ions, $e$ is the electron charge, and $k_e$ is the Coulomb constant.

For $1.05< \alpha < 1.29$, the global minima of $V$ are two sets of inversion-broken square pyramids \cite{ayyadevara2025}, where the apex ion is at $(0,0,\pm z_0)$ and the other four ions form a square at a distance of $z_0/4$ below (above) the $x$-$y$ plane [Fig.~\ref{fig1}(a)]. The height $z_0$ decreases with increasing $\alpha$. The ions freely rotate about the $z$ axis, which blurs the orientation of the four base ions in the fluorescence images due to a finite exposure time. Nonetheless, the inversion breaking is clearly distinguished by the sign of the parity-odd octupole moment, $\psi_{30} \equiv \int d^3 r \, \rho_e(\mathbf{r}) Y_{3,0}(\mathbf{r})$, where $\rho_e(\mathbf{r})$ is the charge distribution and $Y_{3,0}$ is a spherical harmonic. As shown in Figs.~\ref{fig1}(b) and \ref{fig1}(c), we observe stochastic switching between the two polarities, with a dramatic increase in the rate as $\alpha$ is lowered.

\begin{figure}[t]
    \centering
    \includegraphics[width=0.49\textwidth]{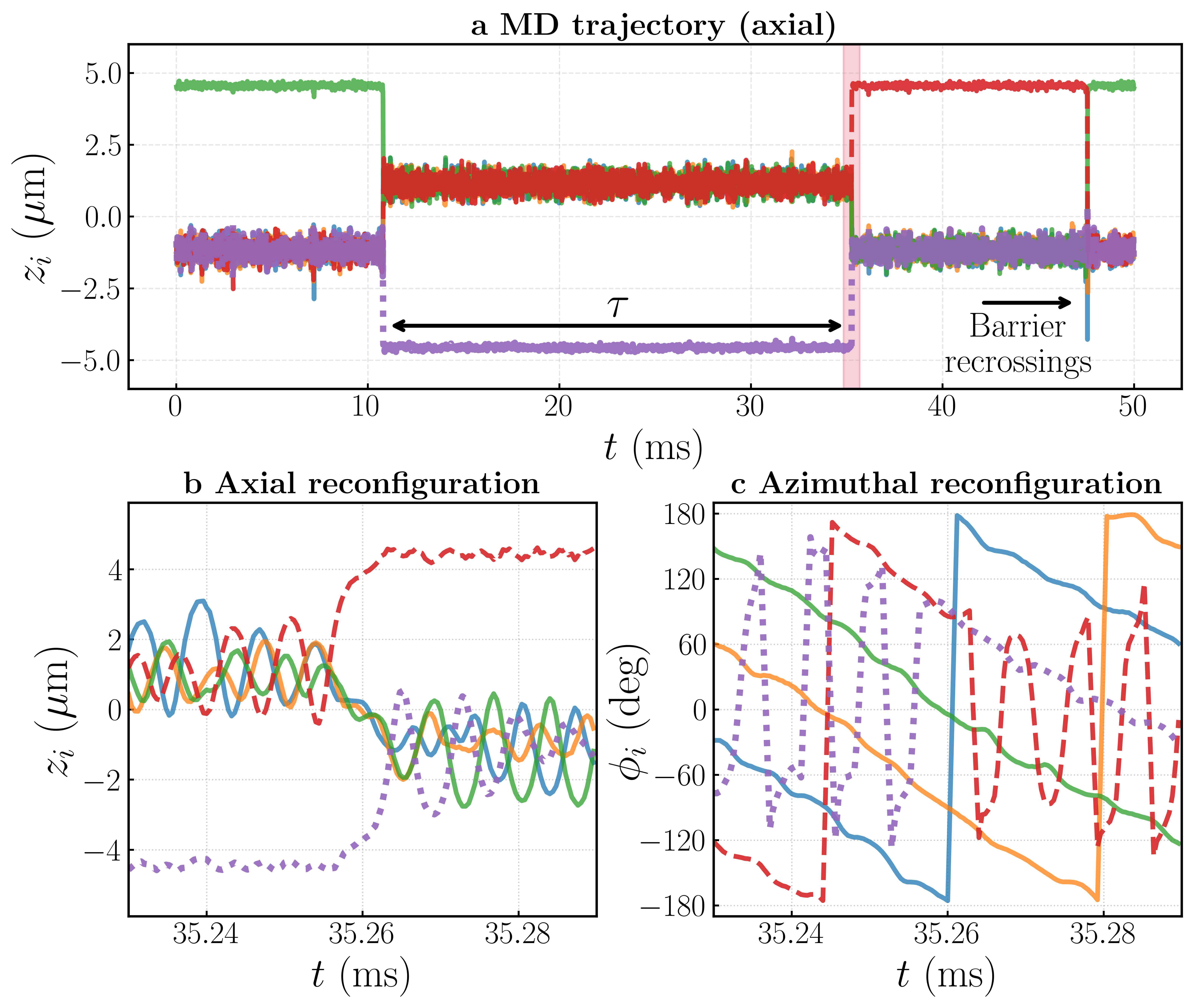}
    \caption{\label{fig2} (\textbf{a}) Trajectory obtained from an MD simulation at experimentally relevant temperature of $2$ mK, showing the \textit{z} coordinates of five $^{40}$Ca$^+$ ions. At the inversion seen at $t \approx 10$~ms, the apex ion (green solid line) moves to the square base, while that base ion (purple dotted line) moves to the inverted apex, completing the \textit{pseudo rotation}. The system remains in this new state for a duration $\tau \approx 25$ ms before inverting back to the initial state, where a different ion (red dashed line) becomes the apex. At $t \approx 48$ ms, we observe rapid barrier recrossings. (\textbf{b, c}) A closer look at cooperative reconfiguration of all five ions during the inversion at $t \approx 35$ ms, capturing the interchange in the roles of apex ions (purple dotted line and red dashed line) and the nearly uniform rotation of the base ions about the $z$ axis. The modulations are due to thermal excitation of the collective modes.}
\end{figure}

{\it \textbf{The pseudo-rotation pathway}}|To identify the inversion pathway and how it changes with $\alpha$, we first note that a switch event occurs over a very short timescale ($\mu$s) during which the cluster rotates uniformly by a small angle $\phi$ about the $z$ axis. This is confirmed by molecular dynamics (MD) simulations discussed below [Fig.~\ref{fig2}(c)]. Hence, the ions invert along the Minimum Energy Path (MEP) in configuration space that connects two pyramids with opposite polarity, whose azimuthal orientations differ by $\phi$. To find this path, it suffices to obtain the MEP corresponding to $\phi=0$ and then superimpose a uniform rotation. We can accurately compute this MEP by employing the climbing-image Nudged Elastic Band (cNEB) method \cite{Henkelman_NEB}, widely used for chemical reactions (see SM \cite{supp} for more details). 

Intuitively, one might expect the inversion to proceed via an {\it umbrella inversion}, where the apex pushes through the center of the square base [Fig.~\ref{fig1}(i)], similar to the case of an ammonia molecule (NH$_3$) \cite{Lehn_1970}. However, this pathway incurs a prohibitively high energy cost due to strong Coulomb repulsion at the planar transition state [Fig.~\ref{fig1}(j)]. Instead, the cluster finds a significantly lower-energy pathway, whereby the apex ion replaces a base ion, while that base ion moves to the new apex [Fig.~\ref{fig1}(e)]. This cooperative exchange is analogous to the \textit{pseudo-rotation} mechanism found in certain molecules, where the displacement of only a few atoms results in an effective rotation of the entire molecule \cite{Berry_1960}. 

Along this pathway, the ions maintain larger pairwise separations compared to the umbrella inversion, greatly reducing the Coulomb energy cost and lowering the activation barrier by approximately two orders of magnitude [Fig.~\ref{fig1}(f)]. In addition, lowering the trap aspect ratio $\alpha$ reduces the activation barrier, since it allows for a larger axial separation $z_0$. This explains the higher inversion rate observed at lower $\alpha$ [Figs.~\ref{fig1}(b) and \ref{fig1}(c)].

{\it \textbf{Isotope substitution}}|The pseudo-rotation pathway is enabled by the permutation symmetry between the apex and the base ions of the pyramid. Breaking this symmetry has dramatic consequences. Experimentally, we break the permutation symmetry by substituting one ion with a heavier isotope ($^{44}\mathrm{Ca}^{+}$; next higher in abundance after $^{40}\mathrm{Ca}^{+}$ \cite{Coplen2002}), which preferentially occupies the apex of the pyramid (see SM \cite{supp}). The pseudo rotation is no longer viable as it does not invert the pyramid. Instead, the ions have to undergo a cooperative exchange of three ions [Fig.~\ref{fig1}(g)] akin to a {\it turnstile rotation} \cite{matsukawa2009pentacoordinate, Couzijn_2010}, with much higher energy barriers [Fig.~\ref{fig1}(h)]. 

The suppression of inversion dynamics due to such high energy barriers is strikingly visible in our real-time traces obtained from fluorescence images. Under identical trapping conditions ($\alpha \approx 1.09$), the $(^{40}\mathrm{Ca}^+)_5$ cluster undergoes rapid inversions [Fig.~\ref{fig1}(c)], whereas the $^{44}\mathrm{Ca}^+(^{40}\mathrm{Ca}^+)_4$ cluster remains locked in one orientation over long periods [Fig.~\ref{fig1}(d)], realizing a structural analog of a giant kinetic isotope effect \cite{Zhao_2017, Petralia_2020}. This observation demonstrates that the rapid inversions indeed occur via the symmetry-enabled pseudo-rotation.

\begin{figure}[t]
    \centering
    \includegraphics[width=0.49\textwidth]{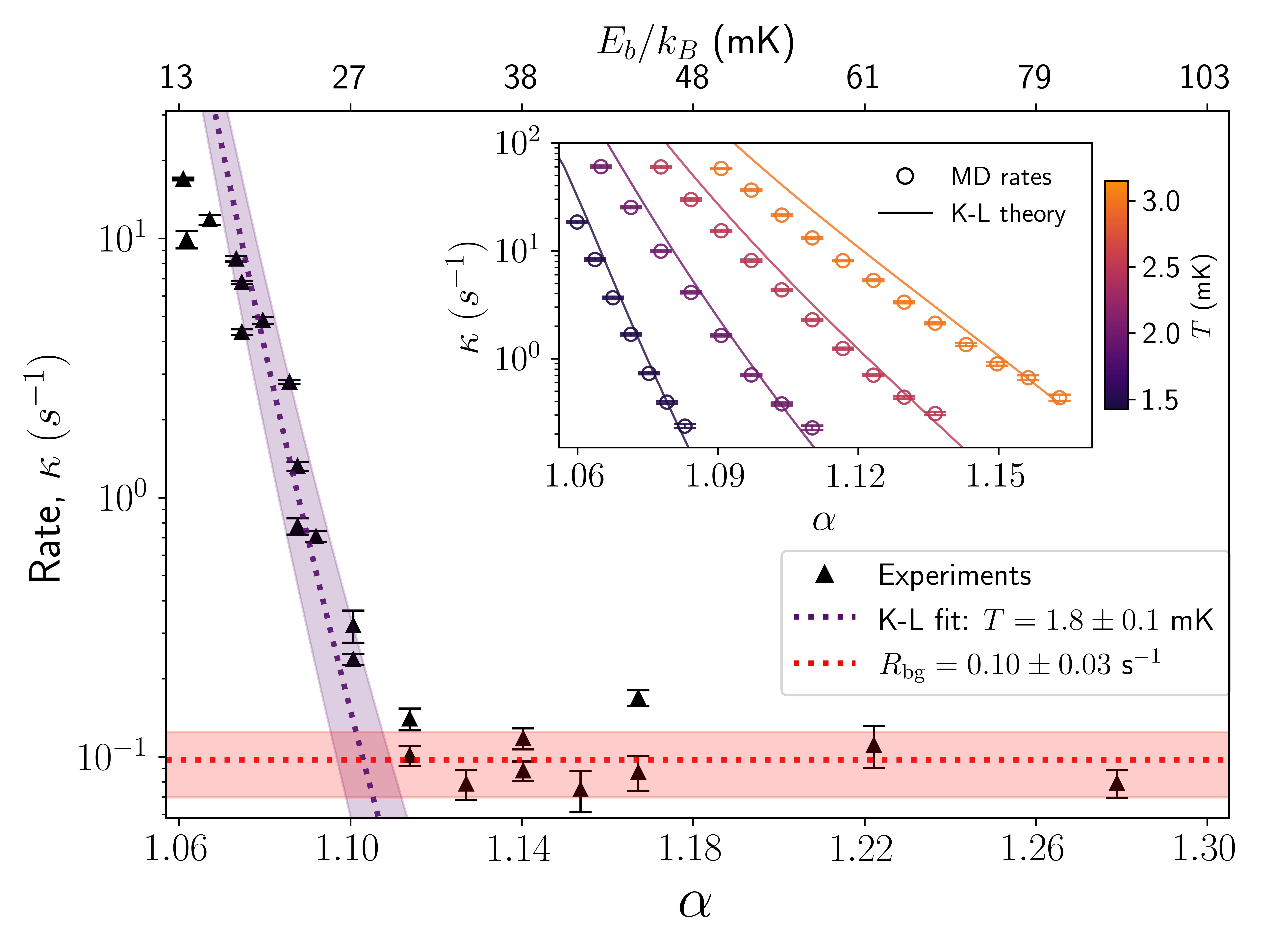}
    \caption{\label{fig3}Experimentally measured pyramidal inversion rates fitted to the K-L theory to determine the cluster temperature $T$ and the inversion rate  $R_{\rm bg}$ due to background collisions. (Inset) Agreement between the inversion rates obtained from MD simulations at four different temperatures to the corresponding K-L theory predictions without any fit parameters.}
\end{figure}

{\it \textbf{Testing Kramers-Langer theory}}|Having confirmed the symmetry-assisted nature of the inversions in the mono-isotopic ($^{40}$Ca$^+$)$_5$ cluster, which can be controlled by tuning the aspect ratio $\alpha$ [Figs.~\ref{fig1}(b) and \ref{fig1}(c)], we turn to the test of the multidimensional K-L theory. First, we compare the K-L theory with MD simulations of the microscopic dynamics at different temperatures and trap aspect ratios to examine its validity. 

The dynamics of the ions are governed by the interplay of the conservative trap and Coulomb potentials, $V(\{\mathbf{r}_i\})$, and the nonconservative forces from the Doppler cooling lasers. In the experiment, the photon scattering rate ($\Gamma/2 \pi \approx 22$ MHz) is much larger than the secular frequencies ($\omega_{x,z}/2 \pi \approx 1$ MHz), so the laser interaction can be modeled by a viscous friction force and a stochastic fluctuating force \cite{Javanainen_1980, Javanainen_1986, Morigi_2001}. Consequently, the motion of the ions is described by the coupled underdamped Langevin equations \cite{Ulm_2013, Pyka_2013}
\begin{equation}
m \ddot{\mathbf{r}}_i + \gamma \dot{\mathbf{r}}_i + \nabla_i V(\{\mathbf{r}_j\}) = \boldsymbol{\xi}_i(t).
\label{eq:langevin}
\end{equation}
Here, $\gamma$ is the friction coefficient determined by the laser detuning and intensity \cite{RevModPhys_Leibfried}, and $\boldsymbol{\xi}_i(t)$ represents a Gaussian white noise satisfying the fluctuation-dissipation relation $\langle \xi_{i,\alpha}(t) \xi_{j,\beta}(t') \rangle = 2 m \gamma k_B T \delta_{ij} \delta_{\alpha\beta} \delta(t-t')$, where $T$ is the equilibrium temperature. 

A typical trajectory obtained from MD simulations of Eq.~\eqref{eq:langevin} is shown in Fig.~\ref{fig2}, where the \textit{pseudo-rotation} mechanism is evident. We extract the inversion rate $\kappa$ from an exponential distribution of the dwell time $\tau$ from more than $2000$ trajectories. We also find barrier recrossings with very short dwell times ($\tau \ll m/\gamma$) typically seen in underdamped Brownian dynamics \cite{recrossings_direct, berezhkovskii2025}, which are excluded from the rate calculation (see SM \cite{supp}).

On the other hand, the escape rate from a local minimum predicted by the multidimensional K-L theory is~\cite{Langer1969}
\begin{equation}
\kappa = \mathcal{N} \frac{\lambda^+}{2\pi} \left( \frac{\det \mathbf{U}}{|\det \mathbf{U}^{\prime}|} \right)^{1/2} \exp(-E_b /k_B T) \, ,
\end{equation}
where $E_b$ is the energy barrier and $\mathbf{U}$, $\mathbf{U}^{\prime}$ are the Hessian matrices at the equilibrium and saddle points along the MEP, respectively. The prefactor $\lambda^+$ is the single positive eigenvalue of the dynamical matrix at the saddle, 
\begin{equation}
    \det \;(m \lambda^2 \mathbf{I} + \gamma \lambda \mathbf{I} + \mathbf{U}^{\prime}) = 0 \, ,
\end{equation}
which accounts for the curvature along the unstable mode at the saddle point and the damping $\gamma$. The additional factor $\mathcal{N}$ denotes the number of equivalent pathways connecting the same initial and final states \cite{Kramers_review_hanggi} (see SM \cite{supp} for an illustration in a toy model).

Before applying this formula to our cluster, we note that the rotational symmetry about the $z$ axis results in a zero-frequency Goldstone mode at both the equilibrium and saddle points \cite{ayyadevara2025}. This free rotation does not affect the inversion dynamics and thus we exclude this mode in the Hessians, as usual \cite{Kramers_review_hanggi}. We also set $\mathcal{N} = 4$ since the apex ion can replace any of the four base ions. As shown in Fig.~\ref{fig3} inset, the rates obtained from MD simulations show excellent parameter-free agreement with the K-L predictions over a wide range of $\alpha$ for temperatures of $1.5$-$3$ mK. The agreement is better at lower temperatures and higher barriers, as expected \cite{Kramers_review_hanggi}.

The experimentally measured inversion rate exhibits a clear threshold behavior (Fig.~\ref{fig3}). For $\alpha < \alpha_{\rm th} \sim 1.1$, it grows rapidly as the barrier height is lowered, whereas for $\alpha > \alpha_{\rm th}$, it saturates to $R_{\text{bg}} = 0.10 \pm 0.03 \, \text{s}^{-1}$ due to background-gas collisions. The excess rate for $\alpha < \alpha_{\rm th}$ follows the K-L prediction over two orders of magnitude for a fitted temperature of $T = 1.8 \pm 0.1$ mK. This agreement confirms that the rapid inversions are indeed thermally activated and validates the applicability of the K-L theory to this strongly coupled, mesoscopic system, allowing us to determine its temperature.

{\it \textbf{Discussion}}|We have tested the multidimensional K-L theory in a genuine many-body setting. By tuning five identical $^{40}\mathrm{Ca}^+$ ions into a bistable square-pyramidal configuration, we uncover a coordinated exchange of the ions enabled by permutation symmetry akin to {\it pseudo rotations} observed in molecules, rather than the conventional {\it umbrella inversion}. The activation barrier can be tuned by varying the trap anisotropy, enabling the experimentally measured inversion rates to span over two decades. The quantitative agreement between our measurements, MD simulations, and the full K-L prediction (up to a small offset due to unavoidable background collisions) not only confirms the theory's predictive power for collective, high-dimensional escape processes but also yields an \textit{in situ} cluster temperature of $1.8 \pm 0.1$ mK, approaching the Doppler-cooling limit.

As the microscopic inversion mechanism itself cannot be directly observed in the experiment, we verify the critical role of permutation symmetry via isotope substitution. By replacing one $^{40}\mathrm{Ca}^+$ with a heavier $^{44}\mathrm{Ca}^+$, we exploit the mass dependence of the harmonic potential to energetically ``pin'' the impurity at the apex. The resulting suppression of thermally activated inversions is analogous to a giant kinetic isotope effect in molecular reactions \cite{Zhao_2017, Petralia_2020}. Interestingly, the residual inversions due to background collisions are also much rarer for the isotope-substituted cluster [see Figs.~\ref{fig1}(d) and \ref{fig3}], which highlights the subtle interplay between the background collisions and the continuous cooling in the rare inversions of bistable ion clusters \cite{Pagano_2019}.

In conclusion, our results present laser-cooled Coulomb clusters as a pristine many-body system for engineering thermally activated dynamics. While we test the Markovian limit of constant Doppler cooling, the effects of different laser cooling protocols \cite{toschek_raman_cooling, rossnagel_thermometry, Neeve_2025} on the inversion dynamics can be investigated. As an alternative to isotope substitution, one can harness internal electronic states or vibrational quanta as a programmable means of breaking the permutation symmetry \cite{RevModPhys_Monroe, Mazzanti_2023, Sun_2024}. This would enable the study of ``active'' symmetry breaking, where the system is driven to specific reaction pathways by coherently preparing the internal states \cite{mallweger2025}, thereby bridging reaction kinetics and quantum control.

\begin{acknowledgments}
\vspace{1em}
{\it Acknowledgments}|We thank Sanjib Sabhapandit and Satya N. Majumdar for fruitful discussions. We acknowledge support from the Department of Science and Technology and the Ministry of Electronics and Information Technology, Government of India, under the “Centre for Excellence in Quantum Technologies” grant with Ref.~No.~4(7)/2020-ITEA. A. Paramekanti acknowledges funding via Discovery Grant RGPIN-2021-03214 from the Natural Sciences and Engineering Research Council (NSERC) of Canada.
\end{acknowledgments}

\onecolumngrid
\clearpage

\begingroup

%\setstretch{1.2}
% Reset figure counter for supplementary material
\setcounter{figure}{0}
\setcounter{section}{0}

% Ensure figures are labeled as S1, S2, etc.
\renewcommand{\thefigure}{SF\arabic{figure}}
\renewcommand{\theHfigure}{SF\arabic{figure}}

\setcounter{equation}{0}
\renewcommand{\theequation}{SE\arabic{equation}}
\renewcommand{\theHequation}{SE\arabic{equation}}

\setcounter{page}{1}
\renewcommand{\thepage}{S\arabic{page}}

\begin{center}
{\LARGE{\textbf{Supplemental Material}}}
\end{center}

\section{Experimental system}

\subsection{Trap potential}

We confine the calcium ions in an end-cap type 3D Paul trap  \cite{schrama_novel_1993} with an rf-drive frequency $\Omega_{\text{rf}} = 2\pi \times 18.26$ MHz \cite{ayyadevara2025}. The time-varying potential ($V_{rf}$) is superimposed with a dc voltage ($U_{dc}$) applied to the inner cylindrical electrodes with a characteristic distance $z_0$. The resulting potential near the trap center is of the form
\begin{equation}
    \Phi(\mathbf{r}, t) = \eta (U_{\text{dc}} + V_{\text{rf}} \cos(\Omega_{\text{rf}} t)) \frac{x^2 + y^2 - 2z^2}{4z_0^2},
\end{equation}
where $\eta$ is the quadrupole efficiency of the trap geometry \cite{schrama_novel_1993}. The motion of a single ion of mass $m$ and charge $Q$ is governed by the Mathieu equation, characterized by the dimensionless parameters $a_u$ and $q_u$ ($u=x,y,z$):
\begin{equation}
    a_z = -2a_x = \frac{4 \eta Q U_{\text{dc}}}{m \Omega_{\text{rf}}^2 z_0^2}, \quad \quad q_z = -2q_x = \frac{-2 \eta Q V_{\text{rf}}}{m \Omega_{\text{rf}}^2 z_0^2}. \label{eq:az_qz}
\end{equation}

We operate the trap in the regime of adiabatic approximation ($|a_u|, q_u^2 \ll 1$), where the ion motion can be separated into a fast micromotion and a slow secular motion. The harmonic secular frequency $\omega_u$ is given by
\begin{equation}
    \omega_u \approx \frac{\Omega_{\text{rf}}}{2} \sqrt{a_u + \frac{q_u^2}{2}}. 
\end{equation}
In the absence of a dc bias ($U_{dc}=0$), $\omega_z=2\omega_x$. A negative dc bias reduces the strength of axial confinement and increases the strength of the transverse confinement simultaneously, enabling the control of the anisotropy $\alpha = \omega_z / \omega_x$.

\begin{figure}[b!]
    \centering
    \includegraphics[width=0.99\linewidth]{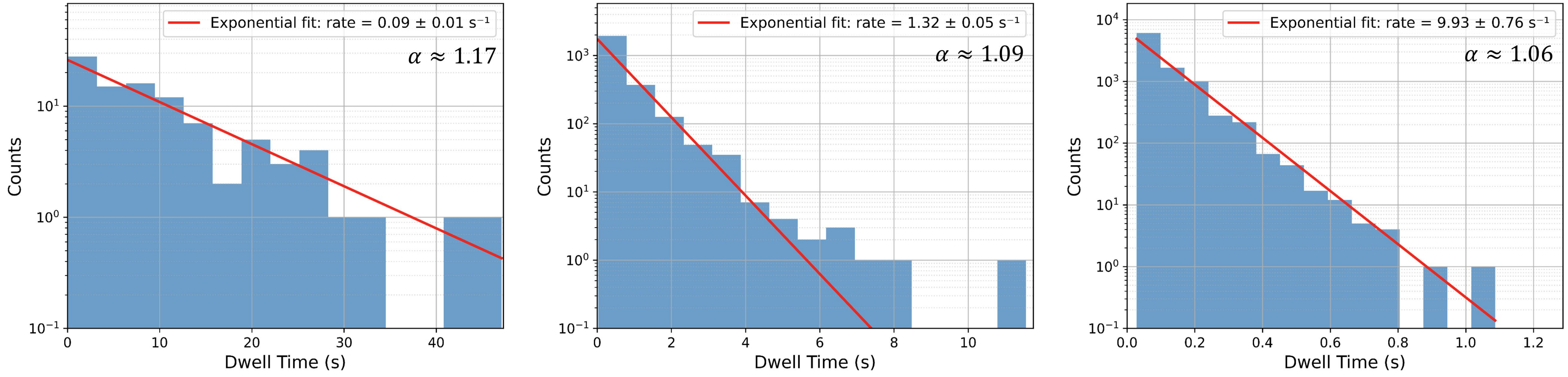}
    \caption{\textbf{Estimation of the experimental inversion rates at different $\alpha$ values.} The dwell times obtained from the time series data are binned into histograms. We obtain the inversion rate and its uncertainty from the exponential fits.}
    \label{fig:expt_rates}
\end{figure}

\subsection{Experimental inversion rates}

To measure the inversion rates experimentally, we use a single Doppler-cooling beam that addresses both radial and axial directions. The cooling laser ($4s$ $S_{1/2} \rightarrow 4p$ $P_{1/2}$  at $ 397$ nm) and the repump laser ($3d$  $D_{3/2} \rightarrow 4p$ $P_{1/2}$ at $ 866$ nm) beams are locked by a wavemeter (HighFinesse WS8-2). The repump beam is locked on the resonance, while the cooling beam is red detuned by $\sim 3 \Gamma$. The intensity of the cooling beam is set to the minimum possible value, while still allowing us to record fluorescence images with an EMCCD (Andor iXon Ultra $897$) at 30 frames per second. For each $\alpha$, the data is recorded for $\sim 60,000$ frames ($\sim30$ minutes). From the $\psi_{30}$ time series, we bin the dwell times into a histogram and fit an exponential to the tail, as shown in Fig.~\ref{fig:expt_rates}.

\section{Numerical calculations}

\subsection{Equilibrium configurations of isotope-substituted cluster}

Since the Mathieu parameters in Eq.~\eqref{eq:az_qz} are mass dependent, the strength of the effective confinement scales with the mass of the trapped species. For the isotope-doped cluster $^{44}\mathrm{Ca}^+(^{40}\mathrm{Ca}^+)_4$, we numerically compute the equilibrium configurations by randomizing initial positions and finding local minima using gradient descent algorithms, shown in Fig.~\ref{fig:ca44_states}. The pentagon global minimum transitions to a pyramid at $\alpha \sim 1.35$, with the heavier isotope $^{40}\mathrm{Ca}^+$ at the apex. A local minimum with lighter isotope $^{40}\mathrm{Ca}^+$ at the pyramid apex only exists for $\alpha > 1.25$, and vanishes from the landscape (signalled by an unstable mode). 

\begin{figure}[t!]
    \centering
    \includegraphics[width=0.725\linewidth]{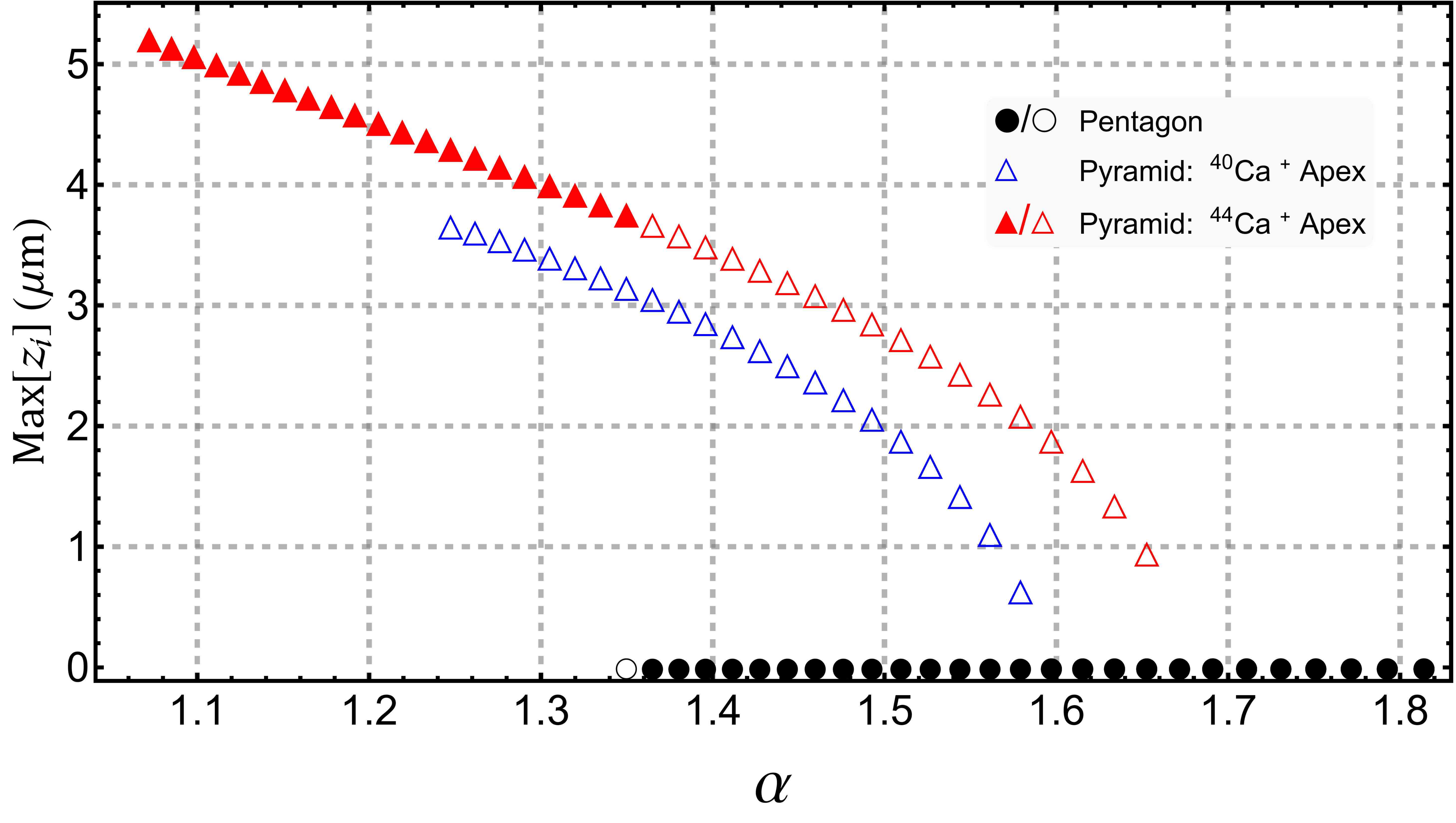}
    \caption{\textbf{Equilibrium configurations of $^{44}\mathrm{Ca}^+(^{40}\mathrm{Ca}^+)_4$ cluster}. The global (local) minima are represented by filled (empty) markers, respectively.}
    \label{fig:ca44_states}
\end{figure}

\subsection{Minimum Energy Pathways}

To identify the transition mechanisms and compute activation barriers, we employ the Climbing Image Nudged Elastic Band (cNEB) method \cite{Henkelman_NEB}. The algorithm discretizes the reaction pathway into a chain of $N$ ``images'' (configurations), $\mathbf{R}_1, \dots, \mathbf{R}_N$, connecting the initial ($\mathbf{R}_{\uparrow}$) and final ($\mathbf{R}_{\downarrow}$) stable states. These images are connected by fictitious springs to maintain spacing (the elastic band), while the potential forces are projected perpendicular to the path tangent (the nudging), allowing the system to relax onto the Minimum Energy Path (MEP). The spring force on the highest-energy image is reversed, driving it to settle exactly at the saddle point.

The relaxation process is sensitive to the initial guess path, particularly in high-dimensional landscapes where multiple saddles may exist. We investigate three distinct initialization scenarios to isolate the relevant reaction coordinates for both identical and isotope-substituted clusters (Fig.~\ref{fig:cneb_path}):

\begin{figure}[tbp]
    \centering
    \includegraphics[width=0.9\linewidth]{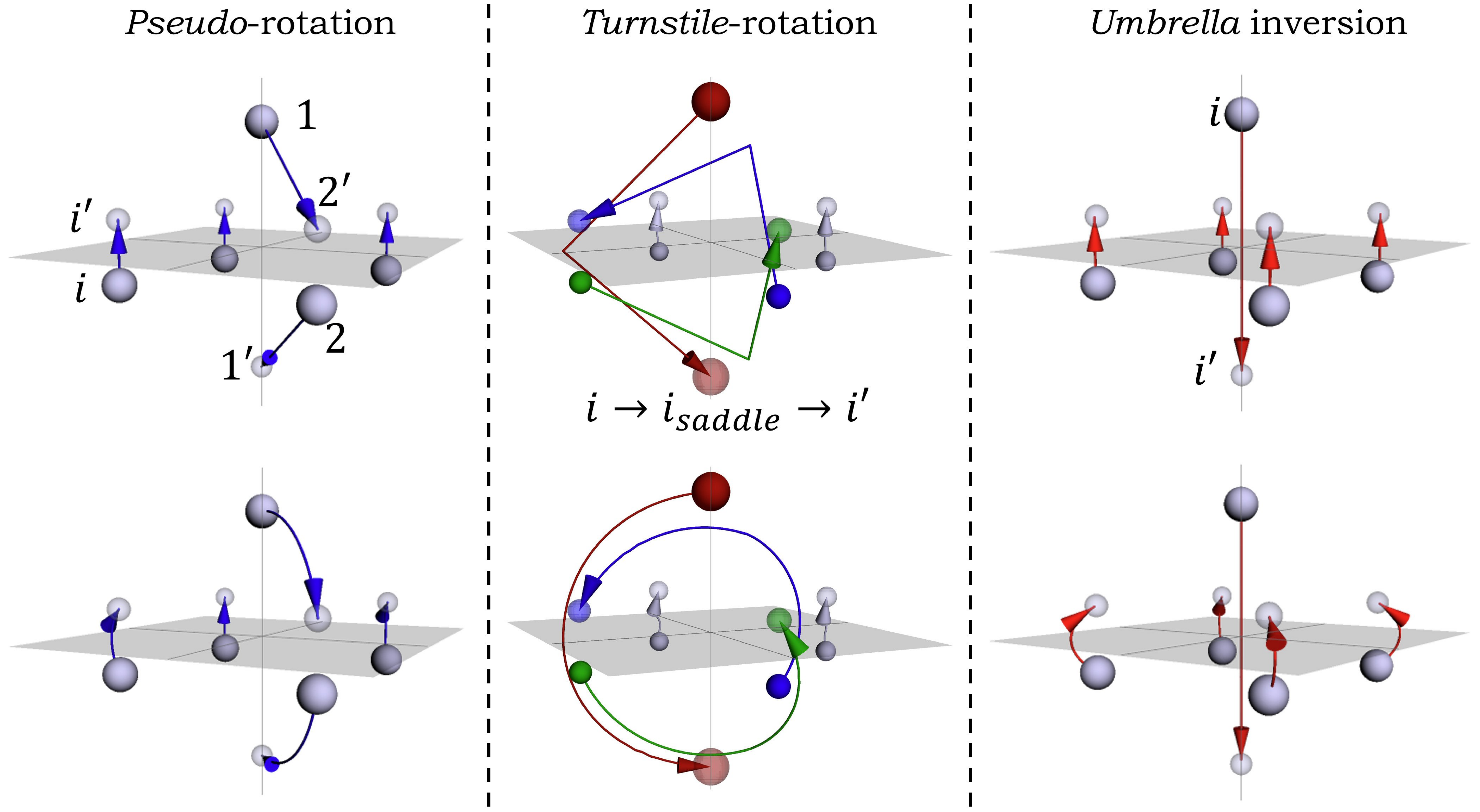}
    \caption{\textbf{Relaxation to the Minimum Energy Path.} (\textbf{Top}) The initial guess pathways used to initialize the cNEB algorithm. (\textbf{Bottom}) The optimized inversion pathways obtained after convergence to the saddle point. From left to right: The symmetry-assisted \textit{pseudo-rotation} (identical ions), the high-barrier \textit{turnstile rotation} (isotope-substituted), and the inaccessible \textit{umbrella inversion}.}
    \label{fig:cneb_path}
\end{figure}

\begin{itemize}

    \item \textbf{Umbrella Inversion:} We define a linear interpolation where ion indices map directly to themselves ($\{ i\rightarrow i'\}$), corresponding to a collective motion through the planar center, as shown in Fig.~\ref{fig:cneb_path}. This path remains dominated by strong Coulomb repulsion, resulting in a high barrier ($> 1000$ mK) for all cluster configurations.

    \item \textbf{Pseudo-Rotation:} For the identical cluster, we initialize a path incorporating permutation symmetry: the apex ion maps to a base ion ($\{ 1\rightarrow 2'\}$), while a base ion maps to the new apex ($\{ 2\rightarrow 1'\}$). This path relaxes to the \textit{Berry pseudo-rotation} MEP, avoiding the high-energy center and reducing the activation barrier by two orders of magnitude (from $\sim 1$ K to $\sim 20$ mK).

     \item \textbf{Turnstile Rotation:} For the isotope-substituted cluster ($^{44}\mathrm{Ca}^+$ at the apex), the pseudo-rotation path is energetically penalized as it requires forcing the heavy impurity into the tighter confinement of the base. Guided by MD simulations, which show a roundabout trajectory, we compute the MEP for a pathway where the heavy apex ion retains its role, while the base ions undergo an internal rotation to achieve inversion. The cNEB optimization converges to a \textit{turnstile rotation} mechanism. The activation barrier is $\sim 270$ mK, which is significantly lower than the umbrella inversion but sufficiently high to suppress thermal activation at experimental temperatures.

\end{itemize}

To verify the convergence of the cNEB pathways, we numerically search for the lowest-energy first-order saddle point using a root-finding algorithm on the force vector ($\nabla V(\mathbf{r}) = 0$). The energy difference between the true saddle point and the cNEB maximum is less than 1 $\mu$K.

\subsection{Estimating Kramers-Langer rates}

While the transition state found by the cNEB method provides the activation barrier, the prefactors for the K-L rate equation are evaluated from the Hessian matrix $\mathbf{H}$ at the equilibrium configuration ($\mathbf{R}_{\uparrow} / \mathbf{R}_{\downarrow}$) and the saddle point ($\mathbf{R}_{\text{saddle}}$). Diagonalizing $\mathbf{H}$ yields the normal mode spectra presented in Table \ref{tab:frequencies}. 

\begin{table}[h!]
    \centering
    \begin{tabular}{c c c}
    \hline \hline
    Mode Index & Equilibrium & Saddle point \\
     & ($\omega_k/\omega_x$)$^2$ & ($\omega_k/\omega_x$)$^2$ \\
    \hline
    1 &  3.14 & 3.15 \\
    2 &  1.78 & 1.85 \\
    3 &  1.78 & 1.75 \\
    ... & ... & ... \\
    13 &  0.03 & 0.04 \\
    14 &  $\mathbf{0.03}$ & $\mathbf{-0.03}$ \\
    15 &  0 & 0 \\
    \hline \hline
    \end{tabular}
    \caption{Calculated eigenvalues of the Hessian matrix (in decreasing order) at the stable square-pyramid and the saddle point for $\alpha \approx 1.09$. The mode $15$ corresponds to the free rotation about the \textit{z}-axis, which is ignored for estimation of K-L rate.}
    \label{tab:frequencies}
\end{table}

The saddle point is characterized by exactly one negative eigenvalue ($\lambda'_1 < 0$), confirming it is a first-order saddle point with the unstable mode along the reaction coordinate. Both configurations share a zero-frequency mode arising from the continuous rotational symmetry about the $z$-axis. As discussed in the main text, this mode is excluded from the determinant ratio in the K-L rate calculation.

\section{Molecular dynamics simulations}

To obtain the theoretical inversion rates, we perform extensive MD simulations by integrating the equations of motion for the 5-ion cluster. The ions are treated as point charges with mass $m$ interacting via the Coulomb force and the external trap potential. The interaction with the laser cooling field is modeled using the Langevin equation:
\begin{equation}
m \ddot{\mathbf{r}}_i = -\nabla_i V(\{\mathbf{r}\}) - \gamma \dot{\mathbf{r}}_i + \boldsymbol{\xi}_i(t),
\end{equation}
where $\gamma$ is the friction coefficient and $\boldsymbol{\xi}_i(t)$ is a stochastic force satisfying the fluctuation-dissipation theorem $\langle \xi_{i\alpha}(t) \xi_{j\beta}(t') \rangle = 2 m \gamma k_B T \delta_{ij} \delta_{\alpha\beta} \delta(t-t')$.

\subsection{Numerical integration}
We use a modified Velocity Verlet integrator adapted for Langevin dynamics \cite{Jensen_2013}. The simulation proceeds in two phases. First, a damping phase ($t < t_{\text{damp}}$) where the system is initialized with random positions and a strong damping force is applied without thermal noise to rapidly settle the ions into the local minimum. Second, a thermal phase where the friction is set to the experimental cooling value ($\gamma / m \approx 6 \times 10^3 \, \text{s}^{-1}$) and the stochastic thermal force is activated. The trajectories are evolved for $10^9$ steps with a time step $\Delta t = 10^{-8}$ s.

\begin{figure}[h]
\centering
\begin{minipage}{0.95\linewidth} 
    \hrule height 1pt \vspace{3pt}
    \textbf{Algorithm 1:} Langevin Dynamics for Ion Inversion
    \vspace{3pt} \hrule \vspace{5pt}
    
    \begin{algorithmic}[1]
        \State \textbf{Initialize:} Random positions $\mathbf{r}$ and velocities $\mathbf{v}$.
        \State \textbf{Parameters:} $\Delta t = 10^{-8}$\,s, $N_{\text{steps}} = 10^9$, \texttt{save\_interval = 100}, Temperature $T$.
        \State \textbf{Define Forces:} $\mathbf{F}_{\text{trap}}(\mathbf{r}) = - \nabla V_{\text{trap}}$, $\mathbf{F}_{\text{coul}}(\mathbf{r}) = - \nabla V_{\text{coul}}$.
        \State
        \For{$step = 1$ to $N_{\text{steps}}$}
            \If{$step < N_{\text{damp}}$} \Comment{Initial Damping}
                \State $\mathbf{F}_{\text{total}} = \mathbf{F}_{\text{trap}} + \mathbf{F}_{\text{coul}} - \gamma_{\text{high}} \mathbf{v}$
                \State Integrate equations of motion (Damped Verlet).
            \Else \Comment{Langevin Dynamics}
                \State Generate Gaussian noise $\boldsymbol{\xi}$ with variance $\sigma^2 = 2 m \gamma k_B T \Delta t$.
                \State $\mathbf{F}_{\text{total}} = \mathbf{F}_{\text{trap}} + \mathbf{F}_{\text{coul}} - \gamma \mathbf{v} + \boldsymbol{\xi}$
                \State Update $\mathbf{r}(t+\Delta t)$ and $\mathbf{v}(t+\Delta t)$.
            \EndIf
            \State
            \If{$step \pmod{\text{save\_interval}} == 0$}
                \State Store $\psi_{30}$ of current configuration.
            \EndIf
        \EndFor
        \State \textbf{Post-Process:} Obtain dwell times $\tau$ from $\psi_{30}$ sign flips. Extract rate from the histogram of dwell times (Fig.~\ref{fig:dwell_times}).
    \end{algorithmic}
    
    \vspace{5pt} \hrule
\end{minipage}
\label{alg:langevin}
\end{figure}

\subsection{Extracting dwell times: barrier recrossings}
We track the octupole moment $\psi_{30}$ of the cluster to identify inversion events, which are flagged by a change in its sign. Due to the low friction (underdamped regime), the system often recrosses the barrier multiple times before settling into the new potential well. These short-time recrossings do not constitute a successful reaction event.

\begin{figure}[tbp]
    \centering
    \includegraphics[width=0.65\linewidth]{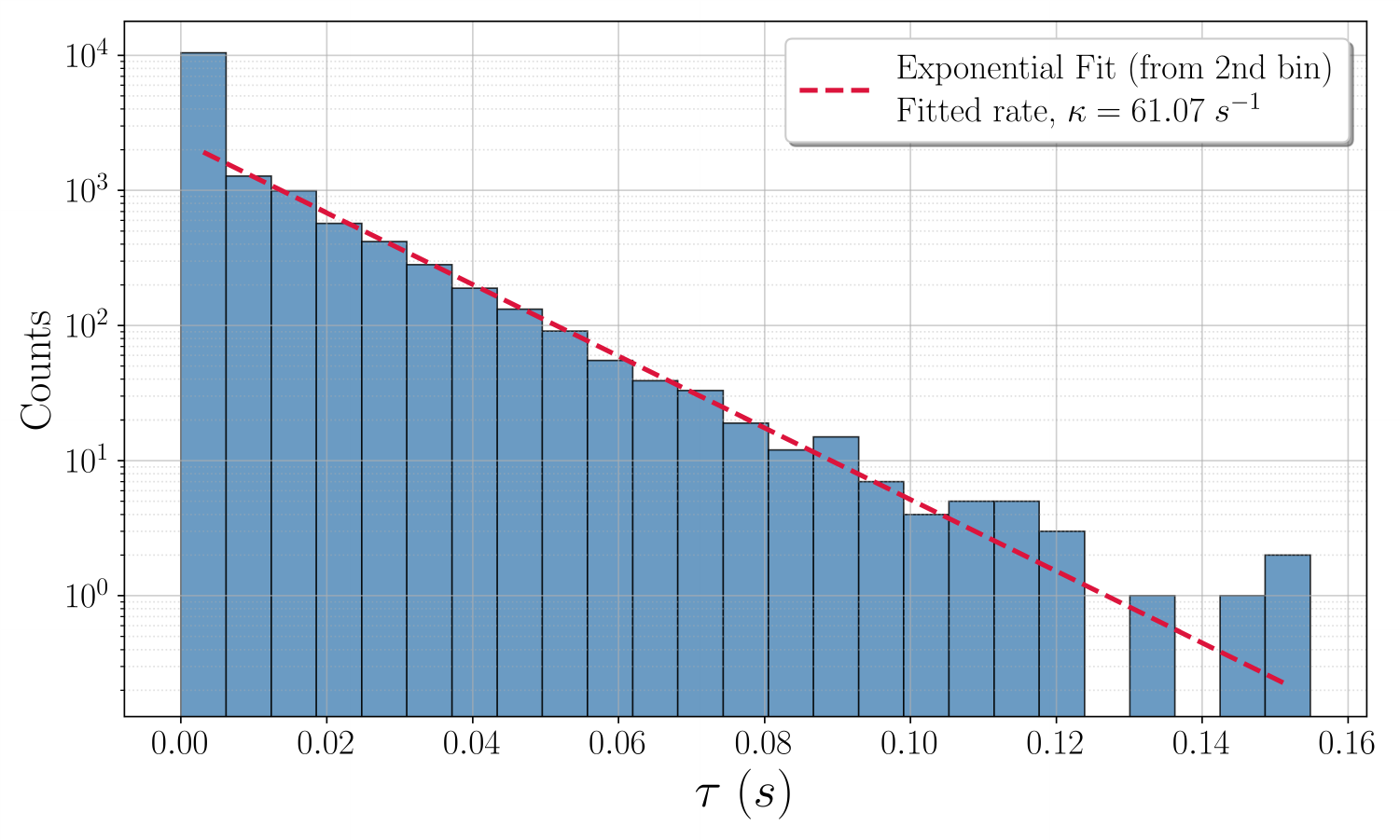}
    \caption{\textbf{Dwell time distribution from MD simulations.} Histogram of time intervals between inversion events from MD simulations. The large count in the first bin corresponds to rapid barrier recrossings (see Fig.~\ref{fig2} in the main text). The rate $\kappa$ is extracted by fitting an exponential decay from the second bin of the histogram ($\tau \gg m/\gamma$).}
    \label{fig:dwell_times}
\end{figure}

To extract the true rate, we analyze the distribution of dwell times between consecutive flips. As shown in Fig. \ref{fig:dwell_times}, there is a clear separation of timescales: a large peak at very short times corresponding to transient recrossings, and an exponential tail corresponding to thermally activated escape. We fit this tail to an exponential distribution $P(\tau) \propto e^{-\kappa \tau}$ to extract the rate $\kappa$, ignoring the recrossing events in the first bin.

\section{Rate expression in presence of degenerate pathways: A 2D toy problem}

The multidimensional Kramers-Langer formula typically calculates the escape rate across a single saddle point. However, certain symmetries of the system can enable multiple, distinct reaction pathways connecting the reactant and product states. A key principle in transition rate theory is that the total rate is the sum of the rates for each independent pathway:
\begin{equation}
    \kappa_{\text{total}} = \sum_{i=1}^\mathcal{N} \kappa_i.
    \label{eq:multiple_paths}
\end{equation}
If all $\mathcal{N}$ pathways are symmetry-equivalent (identical barrier heights $\Delta E$ and Hessian spectra), the total rate simplifies to $\kappa_{\text{total}} = \mathcal{N} \times \kappa_{\text{i}}$. In the case of the 5-ion cluster, the four-fold permutation symmetry implies $\mathcal{N}=4$. 

To verify this multiplicative factor, we perform overdamped Langevin dynamics simulations on a 2D toy model potential designed to have exactly two symmetric saddle points ($\mathcal{N}=2$). The potential is given by:
\begin{equation}
    V(x,y) = 10^4 \left[ (x^2 - 1)^2 + \alpha (x^2 - \beta) y^2 + y^4 \right],
    \label{eq:2d_potential}
\end{equation}
with tunable parameters $\alpha$ and $\beta$. This potential features two stable minima at $(\pm 1, 0)$ separated by a high central barrier at $(0,0)$, forcing the system to escape via one of two symmetric saddles located at $(0, \pm \sqrt{\alpha \beta / 2})$ (see Fig.~\ref{fig:2d-toy-potential}). 

\begin{figure}[b!]
    \centering
    \includegraphics[width=0.95\textwidth]{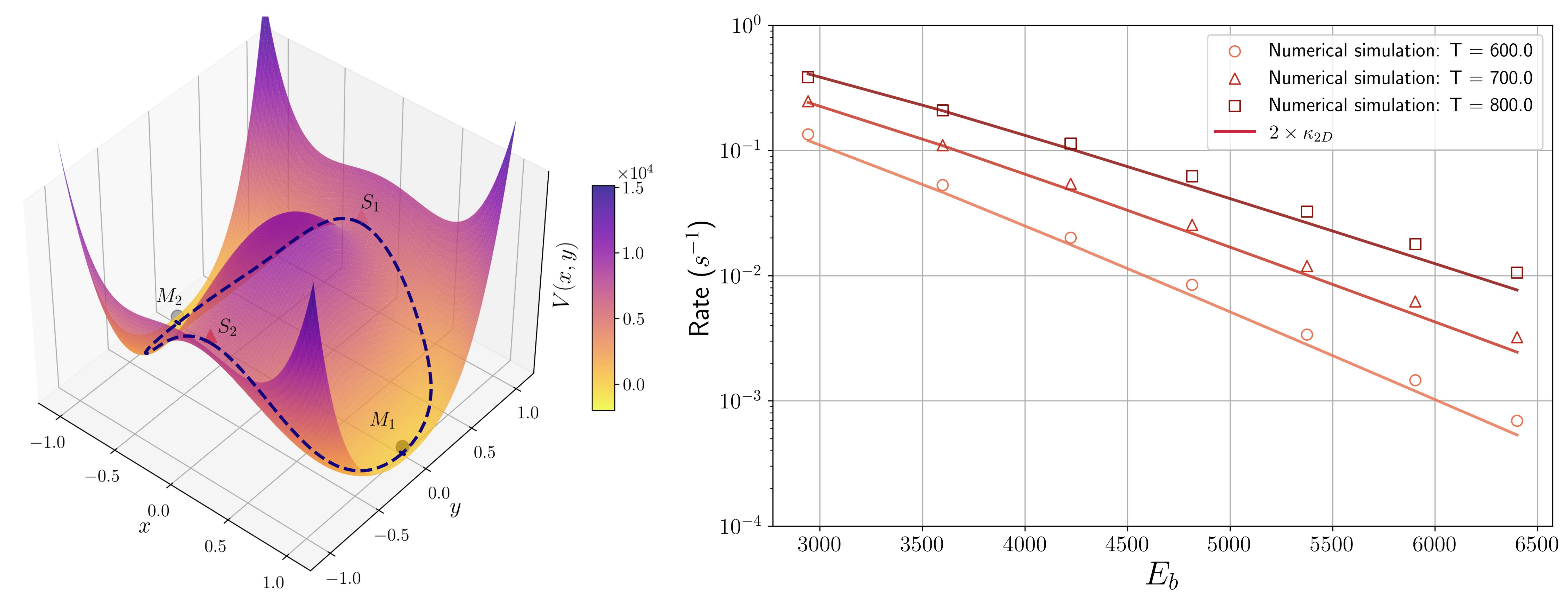}
    \caption{\textbf{Two-Dimensional potential with two-fold pathway degeneracy.} (\textbf{Left}) The potential $V(x,y)$ defined in Eq.~\eqref{eq:2d_potential} features two minima separated by a central maximum at $(0,0)$. Escape from the right well ($M_1$) to the left well ($M_2$) can occur via two symmetry-equivalent saddle points ($S_1$ or $S_2$). (\textbf{Right}) The analytical K-L rates accounting for the number of degenerate saddles $\mathcal{N}=2$ match the numerical results.}
    \label{fig:2d-toy-potential}
\end{figure}

We compute the single-pathway rate $\kappa_{\text{2D}}$ using the Hessian matrices evaluated at one minimum and one saddle point. As shown in Fig.~\ref{fig:2d-toy-potential}, incorporating the degeneracy factor ($\mathcal{N}=2$) yields excellent agreement with the rates extracted from MD simulations across different temperatures. This confirms that as the system fluctuates in the reactant well, it can access any of the symmetry-equivalent saddle points with equal probability. Therefore, the effective flux out of the metastable state is the sum of the fluxes over all degenerate barriers. For the square-pyramidal ion cluster, we use $\mathcal{N}=4$ in the theoretical rate calculation.

\end{document}